\documentclass[journal]{IEEEtran}
\usepackage{amsmath,amssymb,amsthm,color}
\usepackage{algorithm}
\usepackage{algorithmic}
\theoremstyle{definition}
\newtheorem{lem}{Lemma}[section]

\theoremstyle{definition}
\newtheorem{defn}{Definition}[section]

\newtheorem{exmp}{Example}[section]

\theoremstyle{definition}
\newtheorem{thm}{Theorem}

\theoremstyle{remark}
\newtheorem*{rem}{Remark}

\begin{document}
\title{Three Theorems on odd degree Chebyshev polynomials and more generalized permutation 
    polynomials over a ring of module $2^w$}


\author{\IEEEauthorblockN{Atsushi Iwasaki\IEEEauthorrefmark{1},
Ken Umeno\IEEEauthorrefmark{1}}
\IEEEauthorblockA{\IEEEauthorrefmark{1}
Graduate school of Informatics, Kyoto University, Kyoto, Japan}

\thanks{Manuscript received December 1, 2012; revised September 17, 2014. 
Corresponding author: M. Shell (email: http://www.michaelshell.org/contact.html).}}

\markboth{Journal of \LaTeX\ Class Files,~Vol.~13, No.~9, September~2014}%
{Shell \MakeLowercase{\textit{et al.}}: Bare Demo of IEEEtran.cls for Journals}

\IEEEtitleabstractindextext{%
\begin{abstract}
Odd degree Chebyshev polynomials over a ring of modulo $2^w$ have two kinds of period.
One is an ``orbital period".
Odd degree Chebyshev polynomials are bijection over the ring.
Therefore, when an odd degree Chebyshev polynomial iterate affecting a factor of the ring, we can observe an orbit over the ring.
The `` orbital period " is a period of the orbit.
The other is a ``degree period".
It is observed when changing the degree of Chebyshev polynomials with a fixed argument of polynomials.
Both kinds of period have not been completely studied.
In this paper, we clarify completely both of them.
The knowledge about them enables us to efficiently solve degree decision problem of Chebyshev polynomial over the ring, and so a key-exchange protocol with Chebyshev polynomial over the ring is not secure.
In addition, we generalize the discussion and show that a key-exchange protocol with more generalized permutation polynomials which belong to a certain class is not secure.
\end{abstract}

\begin{IEEEkeywords}
Chebyshev polynomial, permutation polynomial, a ring of modulo $2^w$, cryptography, key-exchange
\end{IEEEkeywords}}

\maketitle

\IEEEdisplaynontitleabstractindextext

%
\IEEEpeerreviewmaketitle

\section{Introduction}
\IEEEPARstart{A}{} polynomial is called permutation polynomial over a finite ring $R$ when $f$ is bijection over $R$.
There are many studies about permutation polynomials.
Almost all studies, $R$ is a finite field.

Rivest studied permutation polynomials over a ring of modulo $2^w$ and showed a necessary and sufficient condition on coefficients that given polynomials over the ring belong to the class of permutation polynomials \cite{Rivest}.
Study about permutation polynomials over the ring  is very important because they are compatible with digital computers and  digital signal processors.
They can calculate values of permutation polynomials over the ring faster than over a finite field because 2 power residue operation is practically negligible.
Then, they are in particular expected to be useful for cryptography and pseudo random number generator, and some applications are already proposed \cite{RC6, Umeno-Kim-Hasegawa, Iwasaki-Umeno}.

One of the applications is a key-exchange protocol with odd degree Chebyshev polynomials over a ring of modulo $2^w$ \cite{Umeno}.
Odd degree Chebyshev polynomials are proven to be permutation polynomials over a ring of modulo $2^w$ and they are commutative each other.
Then, the protocol is constructed by replacing the discrete logarithm problem of Diffie-Hellman key-exchange protocol with the degree decision problem of Chebyshev polynomials over the ring.
Thus, the security is related to difficulty of the degree decision problem  over the ring.
If the problem can efficiently be solved, the protocol is not secure.

Although a key-exchange protocol with Chebyshev polynomials over the real-number interval $[-1,1]$ was proposed earlier than that over a ring of modulo $2^w$ \cite{Kocarev-Tasev}, the degree decision problem of Chebyshev polynomials over $[-1,1]$ was solved and so the key-exchange protocol is regarded as being not secure \cite{Bergamo-D'Arco-Santis-Kocarev}. 
However, we cannot directly adapt the method to solve the problem over a ring of modulo $2^w$ because residue operations do not appear in the problem over $[-1,1]$.
It was also shown that the degree decision problem over a ring of modulo $2^w$ can efficiently be solved when the given argument of Chebyshev polynomial is even \cite{Ishii-Yoshimoto}, but the degree decision problem with odd argument has not been solved.

It is conjectured that the difficulty of the problem over a ring of modulo $2^w$ is related to the periodicity of Chebyshev polynomial over the ring.
Here, odd degree Chebyshev polynomials have two kinds of periodicity.
One is `` orbital period", and the other is ``degree period".
Since odd degree Chebyshev polynomials are permutation polynomials over a ring of modulo $2^w$, when an odd degree Chebyshev polynomial iterate affecting a factor of the ring, we can observe an orbit on the ring.
The `` orbital period " is the period of the orbit.
The ``degree period" is observed when changing the degree of Chebyshev polynomials with fixed arguments of polynomials.
Although there are studies about both kinds of period \cite{Ishii,Yoshioka-Dainobu,Iwasaki-Umeno2}, they have not been completely studied so far.

In this paper, we clarify both kinds of the periodicity.
After that, we show that the degree decision problem over a ring of modulo $2^w$ can efficiently be solved even if the given argument of Chebyshev polynomial is odd, and so the key-exchange protocol is not secure.
It takes only $O(w^4)$ times to solve the problem.
The fact does not mean, however, that Chebyshev polynomials are not useful for all the fields in cryptography.

We solved the problem in 2015 \cite{Iwasaki-Umeno3}, and Kawano and  Yoshioka independently solved the problem at almost the same time \cite{Yoshioka-Kawano}.
The solving method proposed in this paper is more general than them.
Although the degree of the Chebyshev polynomial used in the key-exchange protocol is restricted to odd number, the method proposed in this paper can solve the problem even if the solution is even.

In addition, we discuss the reason why the key-exchange protocol with Chebyshev polynomials is not secure and generalize the discussion.
We show that a key exchange protocol with a set of permutation polynomials is not secure if the permutation polynomials in the set satisfy some conditions which odd degree Chebyshev polynomials also satisfy.
If it takes $O\left(f(w)\right)$ times to calculate the value of permutation polynomials for given argument and iteration number, 
it takes only $O\left(w\cdot f(w)\right)$ times to break the key-change protocol with the polynomials.

This paper is constructed as follows:
in Section 2, we introduce Chebyshev polynomials and a degree decision problem over a ring of modulo $2^w$.
In Sections 3 and 4, orbital periodicity of odd degree Chebyshev polynomials  and periodicity of degree are clarified, respectively.
In Section 5, we show an algorithm to solve degree decision problem.
In Section 6, we discuss about more general permutation polynomials including Chebyshev polynomials and show that a key-exchange protocol with the polynomials is not secure.
Finally, we conclude this paper.

\section{Chebyshev Polynomial and degree decision problem}
In this section, we introduce Chebyshev polynomials, their some properties and degree decision problem.

\begin{defn}
Assume that $m$ is an integer.
A Chebyshev polynomial of $m$ degree $T_m(X)$ is defined as a polynomial satisfying
\begin{align*}
T_m(\cos \theta)=\cos m\theta.
\end{align*}  
\end{defn}

For example,
\begin{align*}
T_1(X)&=X,\\
T_2(X)&=2X^2-1,\\
T_3(X)&=4X^3-3X,\\
T_4(X)&=8X^4-8X^2+1,\\
T_5(X)&=16X^5-20X^3+5X.
\end{align*}
By the definition, it is clear that
arbitrary Chebyshev polynomials are commutative such that
\begin{align*}
\forall m,n\in\mathbb{Z},\ \ T_m\circ T_n(X)=T_n\circ T_m(X)=T_{mn}(X).
\end{align*}
It is also clear that the following relation is satisfied.
\begin{align*}
\forall m,n\in\mathbb{Z},\ \ 2T_{m}(X)T_{n}(X)=T_{m+n}(X)+T_{m-n}(X).
\end{align*}

It is known that arbitrary odd degree Chebyshev polynomials are permutation polynomials over a ring of modulo $2^w$ \cite{Umeno}, which means that they are bijective over the ring.

It takes only $O(w^3)$ times to calculate the value of Chebyshev polynomial over a ring of modulo $2^w$.

From the above, the key-exchange protocol with odd degree Chebyshev polynomials was proposed.
The protocol replaced the discrete logarithm problem of Diffie-Hellman key-exchange protocol with the degree decision problem over a ring of modulo $2^w$ \cite{Umeno}. 

\begin{defn}
A degree decision problem over a ring of modulo $2^w$ is as follows:
find $m\in\mathbb{Z}/2^w\mathbb{Z}$ satisfying
\begin{align*}
\Bar{Y}\equiv T_m(\Bar{X}) \mod2^w,
\end{align*}
with given $\Bar{X},\ \Bar{Y}\in\mathbb{Z}/2^w\mathbb{Z}$.
\end{defn}

\section{Orbital period of odd degree Chebyshev polynomials over a ring of modulo $2^w$}

In this section, we prove a theorem about orbital period.
First, we prove some lemmas which are needed for proving the theorem.

\begin{lem}
\label{lem1-1}
Assume that $X_1=(2A_1-1)\cdot2^{k_1}\pm1$ and $X_2=(2A_2-1)\cdot2^{k_2}$, where $A_1$, $A_2$, $k_1$, $k_2$ $\in \mathbb{N}$ and $k_1\geq2$.
For $r\geq2$,
\begin{align}
\label{siki1-1}&T_{2^r}(X_1)\equiv 1\mod 2^{k_1+r+2},\\
\label{siki1-2}&T_{2^r}(X_2)\equiv 1\mod 2^{k_2+r+2}.
\end{align}
\end{lem}

\noindent {\bf Proof}
In the case $r=2$,
\begin{align*}
&T_{2^2}(X_1)\\
=&8X_1^4-8X_1^2+1\\
=&8\{(2A_1-1)\cdot2^{k_1}\pm1\}^4-8\{(2A_1-1)\cdot2^{k_1}\pm1\}^2+1\\
\equiv&1\mod 2^{k_1+2+2},\\
&T_{2^2}(X_2)\\
=&8\{(2A_2-1)\cdot2^{k_2}\}^4-8\{(2A_2-1)\cdot2^{k_2}\}^2+1\\
\equiv&1\mod 2^{k_2+2+2},
\end{align*}
Then, (\ref{siki1-1}) and (\ref{siki1-2}) are true.

Assume that (\ref{siki1-1}) and (\ref{siki1-2}) are true with $r=r_0$.
We will consider the case $r=r_0+1$.
\begin{align*}
&T_{2^{r_0+1}}(X_1)\\
\equiv&2\{T_{2^{r_0}}(X_1)\}^2-1 \mod2^{k_1+r_0+3}\\
\equiv&2\{T_{2^{r_0}}(X_1)\mod 2^{k_1+r_0+3}\}^2-1\mod2^{k_1+r_0+3}\\
\equiv &1\mod2^{k_1+r_0+3},\\
&T_{2^{r_0+1}}(X_2)\\
\equiv&2\{T_{2^{r_0}}(X_2)\}^2-1 \mod2^{k_2+r_0+3}\\
\equiv&2\{T_{2^{r_0}}(X_2)\mod 2^{k_2+r_0+3}\}^2-1\mod2^{k_2+r_0+3}\\
\equiv &1\mod2^{k_2+r_0+3}.
\end{align*}
Then, (\ref{siki1-1}) and (\ref{siki1-2}) are true with $r=r_0+1$.
From the above, the lemma is true.\qed

\begin{lem}
\label{lem1-2}
Assume that $X_1=(2A_1-1)\cdot2^{k_1}\pm1$ and $X_2=(2A_2-1)\cdot2^{k_2}$, where $A_1$, $A_2$, $k_1$, $k_2$ $\in \mathbb{N}$ and $k_1\geq2$.
For $r\geq2$,
\begin{align}
\label{siki2-1}T_{2^r\pm1}(X_1)&\equiv X_1+2^{k_1+r+1}\mod2^{k_1+r+2},\\
\label{siki2-2}T_{2^r\pm1}(X_2)&\equiv X_2+2^{k_2+r}\mod2^{k_2+r+1}.
\end{align}
\end{lem}

\noindent {\bf Proof}
In the case $r=2$,
\begin{align*}
T_{2^2+1}(X)&=16X^5-20X^3+5X,\\
T_{2^2-1}(X)&=4X^3-3X.
\end{align*}
Then, (\ref{siki2-1}) and (\ref{siki2-2}) are true.

Assume that (\ref{siki1-1}) and (\ref{siki1-2}) are true with $r=r_0$.
We will consider the case $r=r_0+1$.
By Lemma \ref{lem1-1}, 
\begin{align*}
&T_{2^{r_0+1}\pm1}(X_1)\\
=&2T_{2^{r_0}\pm1}(X_1)T_{2^{r_0}}(X_1)-X_1\\
\equiv&2\{T_{2^{r_0}\pm1}(X_1)\mod{2^{k_1+r_0+3}}\}\{T_{2^{r_0}}(X_1)\mod{2^{k_1+r_0+3}}\}\\&\ \ \ \ \ \ \ \ \ \ \ \ \ \ \ \ \ \ \ \ \ \ \ \ \ \ \ \ \ \ -X_1\mod{2^{k_1+r_0+3}}\\
\equiv&X_1+2^{k_1+r_0+2}\mod{2^{k_1+r_0+3}},\\
&T_{2^{r_0+1}\pm1}(X_2)\\
=&2T_{2^{r_0}\pm1}(X_2)T_{2^{r_0}}(X_2)-X_2\\
\equiv&2\{T_{2^{r_0}\pm1}(X_2)\mod{2^{k_2+r_0+2}}\}\{T_{2^{r_0}}(X_2)\mod{2^{k_2+r_0+2}}\}\\&\ \ \ \ \ \ \ \ \ \ \ \ \ \ \ \ \ \ \ \ \ \ \ \ \ \ \ \ \ \ -X_2\mod{2^{k_2+r_0+2}}\\
\equiv&X_2+2^{k_2+r_0+1}\mod{2^{k_2+r_0+2}}.
\end{align*}
Then, (\ref{siki2-1}) and (\ref{siki2-2}) are true with $r=r_0+1$. 
From the above, the lemma is true.\qed

\begin{lem}
\label{lem1-3}
Assume that $X_1=(2A_1-1)\cdot2^{k_1}\pm1$ and $X_2=(2A_2-1)\cdot2^{k_2}$, where $A_1$, $A_2$, $k_1$, $k_2$ $\in \mathbb{N}$ and $k_1\geq2$.
For $r\geq2$,
\begin{align}
\label{siki3-1}T_{3\cdot2^r\pm1}(X_1)&\equiv X_1+2^{k_1+r+1}\mod2^{k_1+r+2},\\
\label{siki3-2}T_{3\cdot2^r\pm1}(X_2)&\equiv X_2+2^{k_2+r}\mod2^{k_2+r+1}.
\end{align}
\end{lem}

\noindent {\bf Proof}
The following calculations prove the lemma.
\begin{align*}
&T_{3\cdot2^r\pm1}(X_1)\\
=&2T_{2^{r+1}\pm1}(X_1)T_{2^r}(X_1)-T_{2^r\pm1}(X_1)\\
\equiv&2\{T_{2^{r+1}\pm1}(X_1)\mod{2^{k_1+r+2}}\}\{T_{2^r}(X_1)\mod2^{k_1+r+2}\}\\&\ \ \ \ \ \ \ \ \ \ \ \ \ \ -\{T_{2^r\pm1}(X_1)\mod2^{k_1+r+2}\mod{2^{k_1+r+2}}\\
\equiv&X_1+2^{k_1+r+1}\mod{2^{k_1+r+2}},\\
&T_{3\cdot2^r\pm1}(X_2)\\
=&2T_{2^{r+1}\pm1}(X_2)T_{2^r}(X_2)-T_{2^r\pm1}(X_2)\\
\equiv&2\{T_{2^{r+1}\pm1}(X_2)\mod{2^{k_2+r+1}}\}\{T_{2^r}(X_2)\mod2^{k_2+r+1}\}\\&\ \ \ \ \ \ \ \ \ \ \ \ \ \ -\{T_{2^r\pm1}(X_2)\mod2^{k_2+r+1}\mod{2^{k_2+r+1}}\\
\equiv&X_2+2^{k_2+r}\mod{2^{k_1+r+1}}.
\end{align*}
\qed

\begin{lem}
\label{lem1-4}
Assume that $X_1=(2A_1-1)\cdot2^{k_1}\pm1$ and $X_2=(2A_2-1)\cdot2^{k_2}$, where $A_1$, $A_2$, $k_1$, $k_2$, $B$ $\in \mathbb{N}$ and $k_1\geq2$.
For $r\geq2$,
\begin{align}
\label{siki4-1}T_{(2B-1)\cdot2^r\pm1}(X_1)&\equiv X_1+2^{k_1+r+1}\mod2^{k_1+r+2},\\
\label{siki4-2}T_{(2B-1)\cdot2^r\pm1}(X_2)&\equiv X_2+2^{k_2+r}\mod2^{k_2+r+1},
\end{align}
where $B$ is a natural number.
\end{lem}

\noindent {\bf Proof }
It has been shown that (\ref{siki4-1}) and (\ref{siki4-2}) are true in the cases of $B=1$ and $2$.
We consider the case $B\geq3$.
Assume that (\ref{siki4-1}) and (\ref{siki4-2}) are true at $B=B_0$ and at $B=B_0+1$.
\begin{align*}
&T_{(2B_0+3)\cdot2^r\pm1}(X_1)\\
\equiv&2T_{(2B_0+1)\cdot2^r\pm1}(X_1)T_{2^{r+1}}(X_1)\\&\ \ \ \ \ \ \ \ -T_{(2B_0-1)\cdot2^r\pm1}(X_1) \mod2^{k_1+r+2}\\
\equiv&2\{X_1+2^{k_1+r+2}\}\{1\}-\{X_1+2^{k_1+r+2}\}\\
\equiv&X_1+2^{k_1+r+2},\\
&T_{(2B_0+3)\cdot2^r\pm1}(X_2)\\
\equiv&2T_{(2B_0+1)\cdot2^r\pm1}(X_2)T_{2^{r+1}}(X_2)\\&\ \ \ \ \ \ \ \ -T_{(2B_0-1)\cdot2^r\pm1}(X_2) \mod2^{k_2+r+1}\\
\equiv&2\{X_2+2^{k_2+r+1}\}\{1\}-\{X_2+2^{k_2+r+1}\}\\
\equiv&X_2+2^{k_2+r+1}.
\end{align*}
Then,  (\ref{siki4-1}) and (\ref{siki4-2}) are true at $B=B_0+2$.
From the above, the lemma is true.\qed

\begin{lem}
\label{lem1-5}
Assume that $p$, $m$ and $X_0$ are natural numbers satisfying 
\begin{align*}
T_p(X_0)\equiv X_0+2^m\mod 2^{m+1}.
\end{align*}
If $m\leq w$,

\begin{align*}
&\{T^i_p(X_0)\mod2^w|i=0,1,2,\cdots,2^{w-m}-1\}\\
=&\{X_0+k\cdot2^m\mod2^w | k=0,1,2,\cdots,2^{w-m}-1\}.
\end{align*}
\end{lem}

\noindent Proof of Lemma \ref{lem1-5} is shown in Ref. \cite{Yoshioka-Dainobu}.

\begin{thm}
\label{thm1}
The orbital periods are distributed according to Table \ref{orbital-period}.
\end{thm}
\begin{table}[h]
\begin{center}
\caption{Orbital periods of odd degree Chebyshev polynomials. Here, $A$, $B$, $r$, $k_1$, $k_2$ $\in\mathbb{N}$, $2\leq r\leq w-1$, $2\leq k_1 \leq w-1$ and $k_2\leq w-1$.}
\label{orbital-period}
\begin{tabular}{c c c}
\hline
Initial Point & Degree &Orbital Period \\
\hline
arbitrary & $1$, $2^w-1$ & 1\\
$(2A-1)\cdot2^{k_1}\pm1$ & $(2B-1)\cdot2^r\pm1$ & $\max(2^{w-k_1-r-1}, 1)$\\
$(2A-1)\cdot2^{k_2}$ & $(2B-1)\cdot2^r\pm1$ & $\max(2^{w-k_2-r}, 1)$\\
$0$, $1$, $2^w-1$ & arbitrary & $1$\\
\hline
\end{tabular}
\end{center}
\end{table}

\noindent {\bf Proof }
By Lemmas \ref{lem1-4} and \ref{lem1-5}, it is clear that the second and third lines of Table \ref{orbital-period} are true.

Since $T_1(X)=X$ and $T_{2^w-1}\equiv (X)\mod2^w$, the first line of Table \ref{orbital-period} is true.

Assume that $X_1=1$.
We can express $X_1\equiv (2A_1-1)\cdot2^w+1\mod2^w$. 
Then, by Lemma \ref{lem1-4}, 
\begin{align*}
T_{(2B-1)\cdot2^r\pm1}(X_1)
&\equiv X_1+2^{w+r+1}\mod 2^{w+r+2}\\
&\equiv X_1\mod2^w.
\end{align*}
Similarly, we can get $T_{(2B-1)\cdot2^r\pm1}(0)\equiv0\mod2^w$ and $T_{(2B-1)\cdot2^r\pm1}(2^w-1)\equiv 2^w-1\mod2^w$.
Then, the fourth line of Table \ref{orbital-period} is true.\qed

\begin{exmp}
Let's consider the orbital period with the initial point $X_0=5$ and the degree $p=3$ over a ring of modulo $2^{7}$.
Since $5=2^2+1$ and $3=2^2-1$, the second line of Table \ref{orbital-period} is applied.
The orbital period is calculated as
\begin{align*}
\max(2^{7-2-2-1},1)=4.
\end{align*}
Indeed,
\begin{align*}
T_{3}(5)\ \ &\equiv 101\mod2^7,\\
T_{3}(101)&\equiv 69\ \mod2^7,\\
T_{3}(69)\ &\equiv37\ \mod2^7,\\
T_{3}(37)\ &\equiv 5\ \ \mod2^7.
\end{align*}
Then, the orbital period is surely 4.
\end{exmp}

\section{Degree period of Chebyshev polynomials}

In this section, we prove a theorem about periodicity of degree.
First, we introduce the following some basic lemmas.

\begin{lem}
\label{lem2-1}
\[\forall m\in\mathbb{N},\ T\left(\frac{\alpha+\alpha^{-1}}{2}\right)=\frac{\alpha^m+\alpha^{-m}}{2}.
\]
\end{lem}

\begin{lem}
\label{lem2-2}
Assume that $s$ and $t$ are natural numbers.
Then,
\[a\equiv b\mod 2^s\Rightarrow a^{2^t}\equiv b^{2^t}\mod 2^{s+t}.\]
\end{lem}

\noindent{Proofs of the above two lemmas are shown in Ref. \cite{Ishii}.}

\begin{lem}
\label{lem2-3}
Assume that $X_1=(2A-1)\cdot2^{k_1}\pm1$ and $\alpha=X_1+\sqrt{X_1^2-1}$, where $A$ and $k_1$ are natural numbers and $2\leq k_1\leq w-4$.
Then,
\begin{align}
&\frac{\alpha^{2^{w-k_1-1}}+\alpha^{-2^{w-k_1-1}}}{2}\equiv1\mod2^w,\\
&\frac{\alpha+\alpha^{-1}}{2}\cdot\frac{\alpha^{2^{w-k_1-1}}-\alpha^{-2^{w-k_1-1}}}{2}\equiv0\mod2^w.
\end{align}
\end{lem}

\noindent {\bf Proof}
Since $k_1\geq2$,
\begin{align*}
\alpha^2&=2X_1^2-1+2X_1\sqrt{X_1^2-1}\\
&\equiv1+2X_1\sqrt{X_1^2-1}\mod2^{k_1+2}.
\end{align*}
Assume that $\exists t_0\in\mathbb{N},\ \alpha^{2^{t_0}}\equiv1+2^{t_0}X_1\sqrt{X^2-1}$.
By Lemma \ref{lem2-2},
\begin{align*}
\alpha^{2^{t_0+1}}&\equiv\{1+2^{t_0}X_1\sqrt{X_1^2-1}\}^2\mod2^{k_1+t_0+2}\\
&\equiv1+2^{t_0+1}X_1\sqrt{X^2-1}\mod2^{k_1+t_0+2}.
\end{align*}
Then, $\forall t\in\mathbb{N},\ \alpha^{2^{t}}\equiv1+2^{t}X_1\sqrt{X^2-1}$, and so
\[\alpha^{2^{w-k_1-1}}\equiv1+2^{w-k_1-1}X\sqrt{X^2-1}\mod2^w.
\]
By the same reason,
\[\alpha^{-2^{w-k_1-1}}\equiv1-2^{w-k_1-1}X\sqrt{X^2-1}\mod2^w.
\]
Form the above,
\begin{align*}
&\frac{\alpha^{2^{w-k_1-1}}+\alpha^{-2^{w-k_1-1}}}{2}\equiv1\mod2^w,\\
&\frac{\alpha+\alpha^{-1}}{2}\cdot\frac{\alpha^{2^{w-k_1-1}}-\alpha^{-2^{w-k_1-1}}}{2}\\
\equiv&\sqrt{X_1^2-1}\cdot2^{w-k_1-1}X_1\sqrt{X^2-1}\mod2^w\\
\equiv&0\mod 2^w.
\end{align*}
\qed

\ \\

\begin{lem}
\label{lem2-4}
Assume that $X_2=(2A-1)\cdot2^{k_2}$ and $\alpha=X_2+\sqrt{X_2^2-1}$, where $A$ and $k_2$ are natural numbers satisfying $k_2\leq w-3$.
Then,
\begin{align}
&\frac{\alpha^{2^{w-k_2}}+\alpha^{-2^{w-k_2}}}{2}\equiv1\mod2^w,\\
&\frac{\alpha^{2^{w-k_2}}-\alpha^{-2^{w-k_2}}}{2}\equiv0\mod2^w.
\end{align}
\end{lem}

\noindent{\bf Proof}
Since $X_2=(2A-1)\cdot2^{k_2}$ and $k_2$ is natural number,
\[\alpha^4\equiv1\mod2^{k_2+2}.
\]
Assume that $\exists t_0\in\mathbb{N},\ \alpha^{2^{t_0}}\equiv1\mod2^{k_2+t_0}$.
Then,
\[\alpha^{2^{t_0+1}}\equiv1^2\mod2^{k_2+t_0+1}.\]
Therefore, $\alpha^{2^{w-k_2}}\equiv1\mod2^w$ and $\alpha^{-2^{w-k_2}}\equiv1\mod2^w$.
From the above, 
\begin{align}
&\frac{\alpha^{2^{w-k_2}}+\alpha^{-2^{w-k_2}}}{2}\equiv1\mod2^w,\\
&\frac{\alpha^{2^{w-k_2}}-\alpha^{-2^{w-k_2}}}{2}\equiv0\mod2^w.
\end{align}
\qed

\begin{lem}
\label{lem2-5}
Assume that $p$ is an odd number and $X_1=(2A-1)\cdot2^{k_1}\pm1$, where $A$ and $k_1$ are natural numbers satisfying $k_1\leq w-4$.
Then,
\[T_{p+2^{w-k_1-1}}(X_1)\equiv T_p(X_1)\mod2^w.
\] 
\end{lem}

\noindent{\bf Proof}
Assume that $\alpha=X+\sqrt{X^2-1}$. 
Then, by Lemmas \ref{lem2-1} and \ref{lem2-3},
\begin{align*}
&T_{p+2^{w-k_1-1}}(X_1)\\
=&\frac{\alpha^{p+2^{w-k_1-1}}+\alpha^{-p-2^{w-k_1-1}}}{2}\\
=&\frac{\alpha^p+\alpha^{-p}}{2}\cdot\frac{\alpha^{2^{w-k_1-1}}+\alpha^{2^{w-k_1-1}}}{2}\\
&\ \ \ \ \ \ \ \ \ \ +\frac{\alpha^p-\alpha^{-p}}{2}\cdot\frac{\alpha^{2^{w-k_1-1}}-\alpha^{2^{w-k_1-1}}}{2}\\
&\equiv T_p(X_1)\mod2^w.
\end{align*}
\qed

\begin{lem}
\label{lem2-6}
Assume that $p$ is an odd number and $X_2=(2A-1)\cdot2^{k_2}$, where $A$ and $k_2$ are natural numbers satisfying $k_2\leq w-3$.
Then,
\[T_{p+2^{w-k_2}}(X_2)\equiv T_p(X_2)\mod2^w.
\] 
\end{lem}

\noindent{\bf Proof}
Assume that $\alpha=X+\sqrt{X^2-1}$. 
Then,  by Lemmas \ref{lem2-1} and \ref{lem2-4},
\begin{align*}
&T_{p+2^{w-k_2}}(X_2)\\
=&\frac{\alpha^{p+2^{w-k_2}}+\alpha^{-p-2^{w-k_2}}}{2}\\
=&\frac{\alpha^p+\alpha^{-p}}{2}\cdot\frac{\alpha^{2^{w-k_2}}+\alpha^{2^{w-k_2}}}{2}\\
&\ \ \ \ \ \ \ \ \ \ +\frac{\alpha^p-\alpha^{-p}}{2}\cdot\frac{\alpha^{2^{w-k_2}}-\alpha^{2^{w-k_2}}}{2}\\
&\equiv T_p(X_2)\mod2^w.
\end{align*}
\qed

\begin{thm}
\label{thm2}
The degree periods of odd degree Chebyshev polynomials are distributed according to Table \ref{degree-period}.
\begin{table}[h]
\begin{center}
\caption{Periods of degree. Here, $A$, $k_1$, $k_2$ $\in\mathbb{N}$, $2\leq k_1 \leq w-4$ and $k_2\leq w-3$.}
\label{degree-period}
\begin{tabular}{c c}
\hline
$X$ & Periodicity of Degree\\
\hline
$(2A-1)\cdot2^{k_1}\pm1$& $T_{p+2^{w-k_1-1}}(X)\equiv T_p(X)\mod 2^w$\\
& $T_{p+2^{w-k_1-2}}(X)\not\equiv T_p(X)\mod 2^w$\\
$(2A-1)\cdot2^{k_2}$ & $T_{p+2^{w-k_2}}(X)\equiv T_p(X)\mod 2^w$\\
& $T_{p+2^{w-k_2-1}}(X)\not\equiv T_p(X)\mod 2^w$\\
otherwise & $T_{p+2}(X)\equiv T_p(X)\mod2^w$\\
\hline 
\end{tabular}
\end{center}
\end{table}
\end{thm}

\noindent{\bf Proof}
It has already shown that $T_{p+2^{w-k_1-1}}(X)\equiv T_p(X)\mod 2^w$
for $X=(2A-1)\cdot2^{k_1}\pm1$ and $T_{p+2^{w-k_2}}(X)\equiv T_p(X)\mod 2^w$ for $X=(2A-1)\cdot2^{k_2}$.
First, we consider the case $p\equiv\pm1\mod2^w$ .
Assume that $X_1=(2A-1)\cdot2^{k_1}\pm1$.
By Theorem \ref{thm1}, the orbital period of $T_{\pm1+2^{w-k_1-2}}$ with the initial value $X_1$ is $2$.
Then, 
\[T_{\pm1+2^{w-k_1-2}}(X_1)\not\equiv X_1\mod2^w.\]
On the other hand, $T_{\pm1}(X_1)\equiv X_1\mod2^w$.
Then,
\[T_{p+2^{w-k_1-2}}(X_1)\not\equiv T_p(X_1)\mod2^w.\]

Next, assume that $p=(2B-1)\cdot2^r\pm1$, where $B$ and $r$ are natural numbers  satisfying $2\leq r\leq w-1$.

We consider the case $r\geq w-k_1-1$.
By Theorem \ref{thm1}, the orbital period of $T_p$ with the initial value $X_1$ is $1$.
On the other hand,  since $\exists B^\prime\in\mathbb{N},\ p+2^{w-k_1-1}=(2B^\prime-1)\cdot2^{w-k_1-2}\pm1$, the orbital period of $T_{p+2^{w-k_1-2}}$ with the initial value $X_1$ is $2$.
Then,
\[T_{p+2^{w-k_1-2}}(X_1)\not\equiv T_p(X_1)\mod2^w.\]

We consider the case $r=w-k_1-2$.
The orbital period of $T_p$ with the initial value $X_1$ is $2$.
On the other hand,  since $\exists B^\prime,r^\prime \in\mathbb{N},\ p+2^{w-k_1-1}=(2B^\prime-1)\cdot2^{w-k_1-2+r^\prime}\pm1$, the orbital period of $T_{p+2^{w-k_1-2}}$ with the initial value $X_1$ is $1$.
Then,
\[T_{p+2^{w-k_1-2}}(X_1)\not\equiv T_p(X_1)\mod2^w.\]

We consider the case $2\leq r\leq w-k_1-3$.
Since the orbital period of $T_p$ with the initial value $X_1$ is $2^{w-k_1-r-1}$, 
\[\forall i,j\in\mathbb{Z}/2^{w-k_1-r-1}\mathbb{Z},\ i\ne j\Rightarrow T_p^i(X_1)\not\equiv T_p^j(X_1)\mod2^w.\]
By the definition of Chebyshev polynomials, $\forall i\in\mathbb{N},\ T_p^i(X_1)=T_{p^i}(X_1)$.
If $i$ is an odd number, $\exists B_i\in\mathbb{N},\ p^i=(2B_i-1)\cdot2^r\pm1$.
Then, if $i$ and $j$ are odd numbers and satisfy $i< j\leq 2^{w-k_1-r-2}$,
\[ T_{(2B_i-1)\cdot2^r\pm1}(X_1)\not\equiv T_{(2B_j-1)\cdot2^r\pm1}(X_1)\mod2^w.
\]
\begin{align*}
\{2(B+2^{w-k_1-r-2})-1\}\cdot2^r\pm1=p+2^{w+k_1-1}.
\end{align*}
By Lemma \ref{lem2-5},
\[T_{p+2^{w-k_1-1}}(X_1)\equiv T_p(X_1)\mod2^w.\]
Then,
\[T_{p+2^{w-k_1-2}}(X_1)\not\equiv T_p(X_1)\mod2^w.\]

From the above, for an arbitrary odd number $p$,
\[T_{p+2^{w-k_1-2}}(X_1)\not\equiv T_p(X_1)\mod2^w.\]
By the same way,  for an arbitrary odd number $p$,
\[T_{p+2^{w-k_1-2}}((2A-1)\cdot2^{k_2})\not\equiv T_p((2A-1)\cdot2^{k_2})\mod2^w.\]

Assume that $X_0$ cannot be written as $(2A-1)\cdot2^{k_1}$ nor $(2A-1)\cdot2^{k_2}$.
By Theorem \ref{thm1}, the orbital period of $T_p$ with the initial value $X_0$ is $1$ where $p$ is an arbitrary odd number.
Then,
\[T_{p+2}(X_0)\equiv T_p(X_0)\mod2^w.\]

From the above, the theorem is true.\qed

\begin{exmp}
Let's consider the degree period with $X_0=5$ over a ring of modulo $2^{6}$.
Since $5=2^2+1$, Theorem \ref{thm2} states that
\begin{align*}
T_{p}(X_0)&\equiv T_{p+2^{6-2-1}}(X_0)\mod2^6,\\
T_{p}(X_0)&\not\equiv T_{p+2^{6-2-2}}(X_0)\mod2^6
\end{align*}
for an arbitrary odd number $p$.
Indeed,
\begin{align*}
T_{3}(5)\ \ &\equiv 37\mod2^6,\\
T_{5}(5)\ \ &\equiv 37\mod2^6,\\
T_{7}(5)\ \ &\equiv 5\mod2^6,\\
T_{9}(5)\ \ &\equiv 5\mod2^6,\\
T_{11}(5)\ \ &\equiv 37\mod2^6,\\
T_{13}(5)\ \ &\equiv 37\mod2^6,\\
T_{15}(5)\ \ &\equiv 5\mod2^6,\\
T_{17}(5)\ \ &\equiv 5\mod2^6.\\
\end{align*}
Then, the statement is surely true.
\end{exmp}

\section{Method of solving degree decision problem}

In this section, we show a method to solve a degree decision problem
by using Lemma \ref{lem1-4}, Theorem \ref{thm1} and Theorem \ref{thm2}.

\begin{rem}
Degree decision problem over a ring of modulo $2^w$

Find $p\in\mathbb{Z}/2^w\mathbb{Z}$ satisfying 
\begin{align*}
\Bar{Y}\equiv T_p(\Bar{X})\mod2^w
\end{align*}
with given $\Bar{X}$ and $\Bar{Y}$.
\end{rem}

Assume that $p=p(l)\cdot2^l$ where $p(l)$ is an odd number and $l$ is a non-negative natural number and $\Bar{X}_l=T_{2^l}(X)\mod2^w$.
Then,
\begin{align*}
T_p(\Bar{X})&=T_{p(l)\cdot2^l}(\Bar{X})\\
&=T_{p(l)}\left(T_{2^l}(\Bar{X})\right)\\
&=T_{p(l)}(\Bar{X}_l).
\end{align*}
Then, it is enough to solve the following problem for $l=0,1,\cdots,w-1$.

Find an odd number $p(l)$ satisfying
\begin{align}
\label{RDDP}\Bar{Y}\equiv T_{p(l)}(\Bar{X}_l)\mod 2^w,
\end{align}
where $\Bar{X}_l\equiv T_{2^l}(\Bar{X})$.

We show a method to solve the problem with a fixed $l$.
There are three cases.
\vspace{3mm}

\noindent{\bf Case 1:}
$\Bar{X}_l$ can be expressed as the following form
\begin{align*}
\Bar{X}_l=(2A-1)\cdot2^k\pm1,
\end{align*}
where $A$ and $k$ are natural numbers satisfying $2\leq k\leq w-4$.
In this case, the algorithm to solve the problem is as follows:\vspace{1mm}
\begin{enumerate}
\item If $\Bar{Y}\equiv\Bar{X}_l\mod 2^w$, output $p(l)=1$ and finish this algorithm.
\item Find a natural number $r\geq2$ satisfying
\begin{align*}
\Bar{Y}\equiv \Bar{X}_l+2^{k+r+1}\mod 2^{k+r+2}.
\end{align*}
If $r$ satisfying the condition does not exist, finish this algorithm since any odd number $p(l)$ does not satisfy (\ref{RDDP}).
\item Set $q\leftarrow 2^{r}+1$ and $m\leftarrow k+r+3$.
\item If $Y\not\equiv T_q(\Bar{X})\mod2^m$, $q\leftarrow q+2^{m-k-2}$.
\item If $m\geq w$, output $p(l)=q$ and finish this algorithm.
Else, $m\leftarrow m+1$ and return 4).
\end{enumerate}
\vspace{1mm}

The operation of step 1) is obviously proper.

By the theorem \ref{lem1-4}, if we cannot find $r$ at step 2),
there are the three possible cases: $\Bar{Y}\equiv T_{1}(\Bar{X}_l)\mod2^w$, $\Bar{Y}\equiv T_{2^w-1}(\Bar{X}_l)\mod2^w$
 and $\Bar{Y}\not\equiv T_p(\Bar{X})\mod2^w$ for an arbitrary odd number $p$.
By the theorem \ref{thm1}, $T_1(\Bar{X}_l)\equiv T_{2^w-1}(\Bar{X}_l)\equiv\Bar{X}_l\mod2^w$.
Since the possibility of $\Bar{Y}\equiv\Bar{X}_l\mod2^w$ is removed at step 1),
$\Bar{Y}\not\equiv T_p(\Bar{X})\mod2^w$ for an arbitrary odd number $p$ if we cannot find $r$.
Therefore, the operation of step 2) is proper.

By the theorem \ref{thm2}, if $\Bar{Y}\equiv T_{q}(\Bar{X}_l)\mod2^{m-1}$ and $\Bar{Y}\not\equiv T_{q}(\Bar{X}_l)\mod2^{m}$,
$\Bar{Y}\not\equiv T_{q+2^{m-k-2}}(\Bar{X}_l)\mod2^{m}$.
Then, the operations of step 3)-5) are proper.

From the above, this algorithm is proper.

Since it takes $O(w^3)$ times to calculate the value of $T_q(\Bar{X})\mod 2^w$, this algorithm requires $O(w^4)$ times if there exists an odd number $p(l)$ satisfying (\ref{RDDP}) and $O(w)$ times if there does not exist.
\vspace{3mm}

\noindent{\bf Case 2:}
$\Bar{X}_l$ can be expressed as the following form
\begin{align*}
\Bar{X}_l=(2A-1)\cdot2^k,
\end{align*}
where $A$ and $k$ are natural numbers satisfying $k\leq w-3$.
In this case, the algorithm to solve the problem is as follows:\vspace{1mm}
\begin{enumerate}
\item If $\Bar{Y}\equiv\Bar{X}_l\mod 2^w$, output $p(l)=1$ and finish this algorithm.
\item Find a natural number $r\geq2$ satisfying
\begin{align*}
\Bar{Y}\equiv\Bar{X}_l+2^{k+r}\mod 2^{k+r+1}.
\end{align*}
If $r$ satisfying the condition does not exist, finish this algorithm since any odd number $p(l)$ does not satisfy (\ref{RDDP}).
\item Set $q\leftarrow 2^{r}\pm1$ and $s\leftarrow k+r+2$.
\item If $Y\not\equiv T_q(\Bar{X})\mod2^m$, $q\leftarrow q+2^{m-k-1}$.
\item If $m\geq w$, output $p(l)=q$ and finish this algorithm.
Else, $m\leftarrow m+1$ and return 4).
\end{enumerate}
\vspace{1mm}

By the same way as the case 1, it is shown that this algorithm is proper.
This algorithm also requires $O(w^4)$ times if there exists an odd number $p(l)$ satisfying (\ref{RDDP}) and $O(w)$ times if there does not exist.
\vspace{3mm}

\noindent{\bf Case 3:}
otherwise.
By the theorem \ref{thm2}, $T_p(l)\mod2^w$ is constant for any $p(l)$.
Then, if $\Bar{Y}\equiv\Bar{X}_l$, an arbitrary odd number $p(l)$ satisfies (\ref{RDDP}).
Else, any odd number $p(l)$ does not satisfy (\ref{RDDP}).
\vspace{3mm}

From the above, the original degree decision problem whose domain of searching  degree is not restrict odd numbers can be {efficiently} solved.
Since it takes only $O(w^3)$ times to calculate the value of $\Bar{X}_l$ for each $l$,
the method to solve the problem takes only $O(w^4)$ times if there exists a solution $p$ and $O(w^3)$ times if there does not exist.

\begin{exmp}
Let's find $p$ satisfying
\begin{align*}
865=T_p(7)\mod 2^{11}.
\end{align*}
In this example,
\begin{align*}
w=&11,\\
\Bar{Y}=&865=27\cdot2^5+1.
\end{align*}

First, we consider the case $l=0$.
\begin{align*}
\Bar{X}_0\equiv7=2^3-1.
\end{align*}
This corresponds to the  case 1.
At step 2) of the algorithm, we cannot find $r$ satisfying the condition.
Then, any odd number $p(0)$ is not satisfy (\ref{RDDP}).

Next, we consider the case $l=1$.
\begin{align*}
\Bar{X}_1=T_2(7) \mod2^{11}=97=3\cdot2^5+1.
\end{align*}
This corresponds to the  case 1.
At step 2) of the algorithm, $r=2$ is chosen.
At step 3) we set $q\leftarrow5=2^2+1$ and $m\leftarrow10=5+2+3$.
\begin{align*}
T_{5}(97)\not\equiv865\mod2^{10}.
\end{align*}
Then, $q\leftarrow q+2^3=13$ and $m\leftarrow m+1$.
\begin{align*}
T_{13}(97)\equiv865\mod2^{11}.
\end{align*}
Since $m=11\geq w$, finish this algorithm.

From the above, $p=13\cdot2^1=26$ is a solution of the problem.
\end{exmp}

\section{Generalization}

In this section, we discuss generalization of the former sections.
We consider a broader class of set of permutation polynomials than the set of odd degree Chebyshev polynomials.

There are two direct and essential reasons why the degree decision problem of Chebyshev  polynomials over a ring of modulo $2^w$ is efficiently solved.
One reason is that the periodicity of degree is completely made clear.
The other reason is that Chebyshev polynomials over the ring have some recursive properties.
In general, such recursive properties are common of permutation polynomials over the ring.
There are also two reasons why we could make the periodicity of degree clear.
One reason is that the orbital period is made clear, and the other reason is that the relation between the orbital period and the degree period is known to be the established.
In connection with the later reason, the arbitrary odd degree Chebyshev polynomials over the ring can be expressed as an iteration of third degree Chebyshev polynomials over the ring.

Based on the above, we consider the following set of permutation polynomials: assume that $\{ P_1(X),P_2(X),\cdots,P_n(X)\}$ is a set of permutation polynomials over a ring of modulo $2^w$ and the value $P_i(\Bar{X})$ can be efficiently calculated for given $i$ and $\Bar{X}$.
Then, there exists a permutation polynomial $F(X)$ which satisfies that
\begin{align*}
\forall i,\ \exists j\ \text{s.t.}\ \forall \Bar{X},\ P_i(\Bar{X})\equiv F^j(\Bar{X})\mod 2^w,
\end{align*}
and so the polynomials $P_i(X)$ are {\it commutative} each other. 
$F(X)$ satisfies at least one of the following conditions. 
\begin{itemize}
\item The values of $F^j(\Bar{X})\mod2^w$ can be directly and efficiently calculated for  given $j$ and $\Bar{X}$.
\item We can efficiently find $i$ which satisfies $P_i(X)\equiv F^j(X)\mod2^w$ for given $j$.
\end{itemize}
 In the case of odd degree Chebyshev polynomials, $F(X)$ is $T_3(X)$ and the both conditions are satisfied.
In the case that the later condition is satisfied, we can calculated $F^j(\Bar{X})\mod2^w$ by finding $i$ and calculating $P_i(\Bar{X})\mod2^w$ even if we cannot directly calculate the value.

In this situation, 
we can construct a key-exchange protocol with the set of polynomials.
That replace Chebyshev polynomials at the key-exchange protocol with the polynomials in the set.
The key-exchange protocol, however, is not secure.
 It means that the following problem can efficiently be solved.
\vspace{3mm}

\noindent{\bf Problem}
Find an integer $j$ satisfying
\begin{align*}
\Bar{Y}\equiv F^j(\Bar{X})\mod2^w,
\end{align*}
where $\Bar{Y}$ and $\Bar{X}$ are given integers.
\vspace{3mm}

\begin{thm}
\label{thm3}
The Problem can efficiently be solved.
\end{thm}

\noindent We show an algorithm to solve the problem.
First, we introduce some lemmas.
\begin{lem}
\label{lem5-1}
Assume that $\Bar{Y}$, $\Bar{X}$ and $j$ are integers, $m$ is a non-negative integer and they satisfy
\begin{align*}
\Bar{Y}\equiv F^j(\Bar{X})\mod2^m.
\end{align*}
Then, there is a non-negative integer $l\leq m$ such that
\begin{align*}
\Bar{Y}\equiv F^{j+2^l}(\Bar{X})\mod2^m.
\end{align*}
\end{lem}

\noindent{\bf Proof}
It is clear that the lemma is true in the case $m=0$.
We consider the case $m=1$.
Since $F(X)$ is a permutation polynomials over a ring of modulo $2^w$,
\begin{align*}
F(0)\equiv 0\mod2,\ F(1)\equiv 1\mod2
\end{align*}
or
\begin{align*}
F(0)\equiv 1\mod2,\ F(1)\equiv 0\mod2
\end{align*}
is practical.
Then,
\begin{align*}
F^{j+2}(\Bar{X})\equiv F^j(\Bar{X})\mod2\equiv \Bar{Y}\mod2.
\end{align*}

Assume that $m^\prime$ is a non-negative smaller integer than $m$.
We consider the case that there exists a non-negative number $l^\prime\leq m^\prime$ such that 
$\Bar{Y}\equiv F^{j+2^{l^\prime}}(\Bar{X})\mod2^{m^\prime}$.
In this case,
\begin{align*}
\Bar{Y}\equiv F^{j+2^{l^\prime}}(\Bar{X})+c2^{m^\prime}\mod2^{m^\prime+1},
\end{align*}
where $c\in\{0,1\}$.
\begin{align*}
F^{j+2^{l^\prime+1}}(\Bar{X})
&\equiv F^{2^{l^\prime}}\left(F^{j+2^{l^\prime}}(\Bar{X})\right)\mod2^{m^\prime+1}\\
&\equiv F^{2^{l^\prime}}\left(F^{j}(\Bar{X})+c2^{m^\prime}\right)\mod2^{m^\prime+1}\\
&\equiv F^{2^{l^\prime}}\left(F^{j}(\Bar{X})\right)+c2^{m^\prime}\mod2^{m^\prime+1}\\
&\equiv \Bar{Y} \mod2^{m^\prime+1}.
\end{align*}

From the above, the lemma is true.\qed

\begin{lem}
\label{lem5-2}
Assume that $\Bar{Y}$, $\Bar{X}$ $j$ and $j^\prime$ are integers, $m$ is a non-negative integer and they satisfy
\begin{align*}
\Bar{Y}&\equiv F^j(\Bar{X})\mod2^m,\\
\Bar{Y}&\equiv F^{j^\prime}(\Bar{X})\mod2^{m+1},
\end{align*}
and $l$ is the minimum non-negative integer satisfying
\begin{align*}
\Bar{Y}&\equiv F^{j+2^l}(\Bar{X})\mod2^m.
\end{align*}
Then,
\begin{align*}
j^\prime\equiv j\mod2^l. 
\end{align*}
\end{lem}

\noindent{\bf Proof}
Assume that $j^\prime=j+a2^l+b$ where $a$ is an integer and $b$ is a non-negative integer satisfying $b<2^l$.
Then,
\begin{align*}
F^{j^\prime}(\Bar{X})
=&F^{j+a2^l+b}(\Bar{X})\\
\equiv&F^{j+b}(\Bar{X})\mod2^m.
\end{align*}
Since $2^l$ is the minimum natural number satisfying $F^j(\Bar{X})\equiv F^{j+2^l}(\Bar{X})\mod2^m$, $b=0$.

Then, the lemma is true.\qed

\begin{lem}
\label{lem5-3}
Assume that $\Bar{Y}$, $\Bar{X}$ and $j$ are integers, $m$ is a non-negative integer and they satisfy
\begin{align*}
\Bar{Y}&\equiv F^j(\Bar{X})+2^m\mod2^{m+1},
\end{align*}
and $l$ is the minimum non-negative integer satisfying
\begin{align*}
\Bar{Y}&\equiv F^{j+2^l}(\Bar{X})\mod2^m.
\end{align*}
If there exists an integer $j^\prime$ such that
$\Bar{Y}\equiv F^{j^\prime}(\Bar{X})\mod2^{m+1}$,
\begin{align*}
\Bar{Y}\equiv F^{j+2^l}(\Bar{X})\mod2^{m+1}.
\end{align*}
\end{lem}

\noindent{\bf Proof}
Assume that 
\begin{align*}
\Bar{Y}&\equiv F^{j+2^l}(\Bar{X})+2^m\mod2^{m+1}.
\end{align*}
For arbitrary integer $a$, 
\begin{align*}
F^{a2^l}(\Bar{Y})
&\equiv F^{a2^l}\left(F^{j+2^l}(\Bar{X})+2^m\right)\mod2^{m+1}\\
&\equiv F^{a2^l}\left(F^{j+2^l}(\Bar{X})\right)+2^m\mod2^{m+1}\\
&\equiv F^{(a+1)2^l}\left(F^{j}(\Bar{X})\right)+2^m\mod2^{m+1}\\
&\equiv F^{(a+1)2^l}\left(F^{j}(\Bar{X})+2^m\right)\mod2^{m+1}\\
&\equiv F^{(a+1)2^l}\left(\Bar{Y}\right)\mod2^{m+1}.
\end{align*}
Then, 
\begin{align*}
F^{j+a2^l}(\Bar{X})
&\equiv F^{a2^l}(\Bar{Y}+2^m)\mod2^{m+1}\\
&\equiv F^{a2^l}(\Bar{Y})+2^m\mod2^{m+1}\\
&\equiv F^{2^l}(\Bar{Y})+2^m\mod2^{m+1}\\
&\equiv F^{2^l}\left(F^{j}(\Bar{X})+2^m\right)+2^m\mod2^{m+1}\\
&\equiv F^{2^l}\left(F^{j}(\Bar{X})\right)\mod2^{m+1}\\
&\equiv F^{j+2^l}(\Bar{X})\mod2^{m+1}\\
&\equiv \Bar{Y}+2^m\mod2^{m+1}.
\end{align*}
By Lemma \ref{lem5-2}, if there exists an integer $j^\prime$ such that $\Bar{Y}\equiv F^{j^\prime}(\Bar{X})\mod2^{m+1}$, there exists an integer $b$ such that
\begin{align*}
\Bar{Y}\equiv F^{j+b2^l}(\Bar{X})\mod2^w.
\end{align*} 

From the above, this lemma is true.\qed
\vspace{3mm}

By using the above lemmas, the problem can be solved.
We propose the following algorithm to solve the problem:
\vspace{1mm}
\begin{enumerate}
\item Set $j\leftarrow0$ and $l\leftarrow0$.
\item If $\Bar{Y}\equiv F^j(\Bar{X})\mod 2^w$, output $j$ and finish this algorithm.
Else, find $m_j$ such that
\begin{align*}
\Bar{Y}\equiv F^j(\Bar{X})+2^{m_j}\mod2^{m_j+1}.
\end{align*}
\item If $\Bar{Y}\equiv F^{j+2^l}(\Bar{X})\mod 2^{m_j+1}$, $j\leftarrow j+2^l$ and return to 2).
\item If $l=w-1$, finish this algorithm. (In this case, any integer $j$ dose not satisfy $\Bar{Y}\equiv F^j(\Bar{X})\mod2^w$.)
Else, $l\leftarrow l+1$ and return to 3).
\end{enumerate}
\vspace{1mm}
If it takes $O\left(f(w)\right)$ times to calculate $F^j(\Bar{X})\mod2^w$ for given $\Bar{X}$ and $j$, this algorithm requires only $O\left(w\cdot f(w)\right)$ times at the worst case.

Then, Theorem \ref{thm3} is proven.

\section{Conclusion}
We completely clarified the orbital periods and degree periods of odd degree Chebyshev polynomials over a ring of modulo $2^w$.
Both of them show a recursive property, and it is shown here that we can efficiently solve a degree decision problem by using the property.
The proven fact here shows that the proposed key-exchange protocol with Chebyshev polynomials is not secure. 
It is also shown that a key-exchange protocol with more generalized permutation polynomials including Chebyshev polynomials is not secure.
However, it does not mean that permutation polynomials over a ring of modulo $2^w$ including Chebyshev polynomials will not be applied to other fields such as symmetric cipher.
The characteristics shown in here would provide useful clues for investigating further applications based on Chebyshev polynomials and other permutation polynomials over a ring of modulo $2^w$.

\end{document}